# Large Seebeck coefficients in Iron-oxypnictides : a new route towards n-type thermoelectric materials

**Loreynne Pinsard-Gaudart[1], David Berardan[1], Julien Bobroff[2], Nita Dragoe[*,1]**

[1] Institut de Chimie Moléculaire et des Matériaux d'Orsay (UMR CNRS 8182), Bât. 410, Univ. Paris-Sud 11, F-91405 Orsay, France.
2 Laboratoire de Physique des Solides (UMR 8502), Bât. 510, Univ. Paris-Sud 11, F-91405 Orsay, France.

Corresponding author: e-mail nita.dragoe@u-psud.fr

Abstract : The iron-oxypnictide compounds, recently reported as a new class of superconductors when appropriately doped, exhibit large Seebeck coefficients, of the order of -100 µV/K, while keeping good electrical conductivity. Their power factor shows a peak at low temperatures, suggesting possible applications of these materials in thermoelectric cooling modules in the liquid nitrogen temperature range.

**Introduction** Rare-earth (RE) iron arsenide oxides (REFeAsO) were initially reported by Quebe et al [1]. Single crystal analysis showed that these compounds crystallize in a P$4/nmm$ structure, containing two REFeAsO units per unit cell, which consists of alternating iron arsenide and rare-earth oxide layers. Little work was published on these materials until it was found recently that, when doped with fluorine on the oxygen site, electron-doped LaAsFeO becomes a superconductor (SC) with $T_c$=26K [2]. Following these surprising findings, it has been noted that the superconductivity is induced as well by hole doping [3]. By adjusting the lattice volume *via* rare-earth replacement, higher critical temperatures were reported [4]. While a huge effort is presently devoted on experimental and theoretical aspects, no clear understanding of both superconducting and normal state properties is achieved yet. In this report, we present evidence of large Seebeck coefficient in these materials coupled with good electrical conductivity. This leads to promising thermoelectric power factors that could enable the use of these materials in thermoelectric modules in refrigeration applications around liquid nitrogen temperatures.

**Experimental** REAsFeO compounds were synthesized, as originally reported, in closed silica tubes heated at high temperatures, backfilled with pure Ar at about 200 mbar pressure (at ambient temperature). We will discuss in the following the synthesis of the compound SmAsFeO$_{0.8}$F$_{0.2}$ (SmAsFeO(F)); other compounds with Nd or La were synthesized in a similar manner and some illustrative results are shown here. SmAs alloys were obtained from Sm metal and As chips following a procedure suggested by Hiscocks and Mullin [5]. The obtained SmAs alloy was mixed with Fe$_2$O$_3$, Fe and with FeF$_3$ in the stoichiometric ratios. The powdered compounds were thoroughly mixed and pressed as small bars. Heating of these bars wrapped in tantalum foil, sealed in silica tubes, at 1150 - 1160°C for more than 40 hours, yielded well crystallized samples. X-ray diffraction patterns were obtained with a Panalytical X'Pert with a Ge(111) incident monochromator and an X'celerator detector. Data were refined by the Rietveld method, by using the GSAS package [6]. Electrical resistivity was measured using a DC four wires method in a closed cycle cryostat, from 300 K to about 30K. The thermoelectric power was measured by a differential method by using the slope of the ΔV versus ΔT curve with gradients up to about 0.2 K/mm, by using a laboratory made system. All transport measurements were performed in a direction perpendicular to the pressing direction. Magnetic properties were measured in a commercial Quantum Design MPMS, under various fields, down to 2K.

**Results and discussion** We report here large thermoelectric power factors, for some of the compounds studied in this family, that indicate potential applications for these materials in thermoelectric cooling modules.

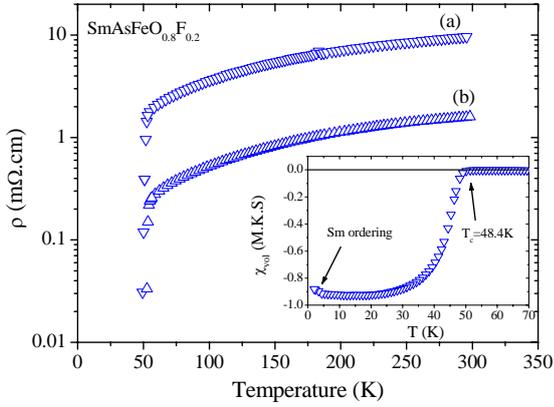

**Figure 1** Temperature dependence of the electrical resistivity of SmFeAsO$_{0.8}$F$_{0.2}$ compounds: first heating (a), second heating (b). Inset: Magnetic susceptibility measured after Zero Field Cooling at H=5G is plotted for the SmFeAsO$_{0.8}$F$_{0.2}$ first heating.

X-ray powder diffraction on a SmAsFeO(F) sample obtained after one heat treatment evidences, besides the P4/nmm peaks, the presence of minor impurities, SmOF and SmAs, estimated at under 10%. Bulk magnetic susceptibility measured after Zero Field Cooling (ZFC) in H=5G shows a sharp superconducting transition with $T_c$=48K as plotted in inset of fig.1. The superconducting fraction of about 90% demonstrates the quality of the SmAsFeO phase in this sample. DC resistivity (Figure 1) and Seebeck coefficient (Figure 2a) confirmed this behaviour, showing a clear SC transition at about 50K with a resistivity of about $2 \cdot 10^{-3}$ Ohm·cm before the transition. A second thermal treatment leads to a slight improvement of the critical temperature and a significant decrease of the resistivity, by one order of magnitude. This treatment induced a slight decrease of the cell volume together with a decrease of the SmOF impurity fraction. These findings are consistent with an increase of the fluorine concentration in the SmAsFeO(F).

The Seebeck coefficient measured after the first heat treatment reaches a maximum value of -90 µV·K$^{-1}$ at about 80K, with a well defined peak. The second thermal treatment decreases this value to about -60 µV·K$^{-1}$ probably following the increase of the electron concentration with doping of fluorine on the oxygen site.

We also obtained large Seebeck coefficients for Nd samples (-100 µV·K$^{-1}$ at 90K) and for La samples (-130 µV·K$^{-1}$). A similar value was reported for a LaFeAsO(F) sample [7]. Up to now, the best thermoelectric performance in the 80K-100K temperature range are found in single crystalline Bi-Sb alloys, with a thermoelectric figure of merit ZT reaching about 0.5 (ZT=S²T/λρ with S the Seebeck coefficient, λ the thermal conductivity and ρ the electrical resistivity). However, these single crystals are both hard to produce and quite brittle, which make them unsuitable for practical applications. Therefore, our results should rather be compared to those obtained in polycrystalline Bi-Sb alloys. As it can be seen in figure 2a, the order or magnitude of the Seebeck coefficient of iron-oxypnictides is very similar to the best values obtained in Bi-Sb alloys [8], and reaches about -130 µV·K$^{-1}$ around 100K. These large Seebeck coefficient values might be related to electron correlations or to a 2D electron confinement in FeAs layers.

Concerning the electrical resistivity, it is moderately higher in iron-oxypnictides than in Bi-Sb alloys, leading to a lower thermoelectric power factor S²/ρ (figure 2b): the power factor reaches 1 mW·m$^{-1}$·K$^{-2}$ in SmFeAsO$_{0.8}$F$_{0.2}$ with is about 7 times lower than in Bi$_{88}$Sb$_{12}$.

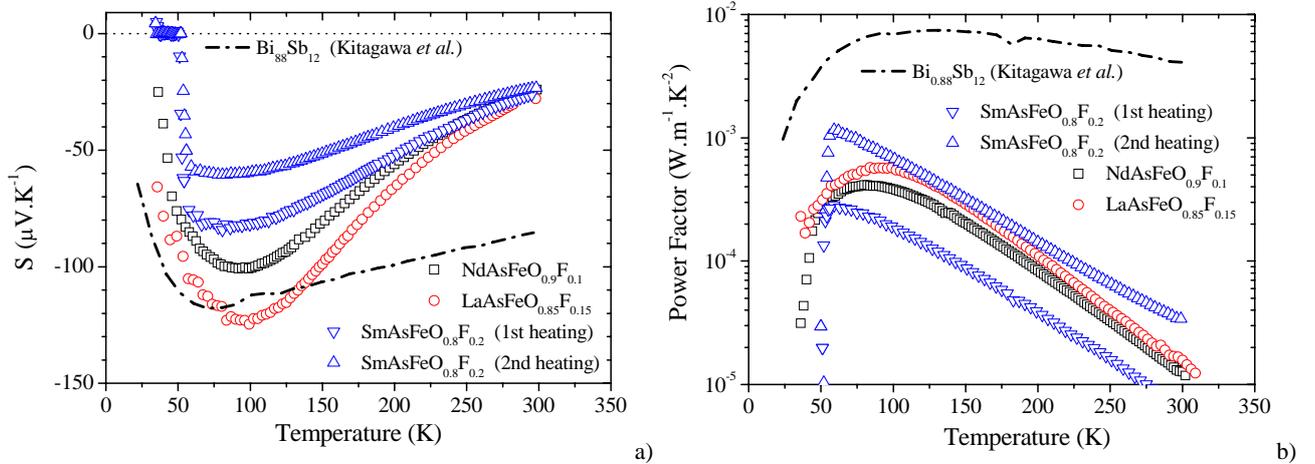

**Figure 2** Temperature dependence of a) the Seebeck coefficient and b) thermoelectric power factor of some REFeAsO$_{1-x}$F$_x$ compounds. The dashed-dotted lines correspond to the data of Bi$_{88}$Sb$_{12}$ reported in ref 9.

However, it must be emphasized that the power factor measured in these compounds is not so far from that of Bi-Sb alloys, as our samples are poorly densified and that much lower electrical resistivity values have been reported for single crystals. This should lead to much higher power factor values. For instance, a resistivity of $\rho \sim 4 \cdot 10^{-4}$ Ohm·cm has been reported at 100K in single crystals of Nd compounds [9], and assuming similar Seebeck coefficient, this would correspond to a power factor of 2.5 mW·m$^{-1}$·K$^{-2}$ which is about the same magnitude as Bi$_{88}$Sb$_{12}$. On the other hand, in the same REFeAsO family, it has been reported that hole doping on the rare-earth oxide layer can yield *p*-type samples [3]. This would be of fundamental interest for applications, since the thermoelectric properties of the known *p*-type compounds are rather poor at low temperature. Therefore, the iron-oxypnictide materials appear to be very promising for thermoelectric cooling applications in the liquid nitrogen temperature range.